\documentclass{aa}  

\usepackage{graphicx}
\usepackage{txfonts}
\usepackage{hyperref}
\hypersetup{
        colorlinks   = true, 
        urlcolor     = black, 
        linkcolor    = blue, 
        citecolor   = blue 
}
\usepackage[all]{hypcap}
\makeatother
\usepackage{txfonts}
\usepackage[colorinlistoftodos]{todonotes}
\usepackage{url}
\usepackage{xcolor}
\usepackage{multicol, blindtext}
\usepackage{txfonts}
\usepackage[Symbol]{upgreek}
\usepackage{subfig}
\usepackage{float}
\usepackage{textgreek}
\usepackage[colorinlistoftodos]{todonotes}

\usepackage{etoolbox}
\usepackage{txfonts}
\usepackage[colorinlistoftodos]{todonotes}
\usepackage{url}
\usepackage{xcolor}

\newcommand{\teff}{T_\mathrm{eff}}
\newcommand{\tint}{T_\mathrm{int}}
\newcommand{\teq}{T_\mathrm{eq}}
\newcommand{\logg}{\mathrm{log}\,g}

\begin{document} 
\title{A dayside thermal inversion in the atmosphere of WASP-19b}

 \titlerunning{Atmosphere of WASP-19b}
 \author{A. S. Rajpurohit \inst{1}, F. Allard \inst{2}, D. Homeier \inst{3}, O. Mousis \inst{4}, S. Rajpurohit \inst{5}}
\institute{Astronomy \& Astrophysics Division, Physical Research Laboratory, Navrangapura, Ahmedabad 380009, India\\
\email{arvindr@prl.res.in}
\and
Univ Lyon, Ens de Lyon, Univ Lyon1, CNRS, Centre de Recherche Astrophysique de Lyon UMR5574, F-69007, Lyon, France
\and
F\"{o}rderkreis Planetarium G\"{o}ttingen  e.V., Nordh\"{a}user Weg 18, 37085, G\"{o}ttingen, Germany
\and
Aix Marseille Universit\'{e}, CNRS, LAM (Laboratoire d'Astrophysique de Marseille) UMR 7326, 13388, Marseille, France 
\and
The Molecular Foundry, Lawrence Berkeley National Laboratory, Berkeley, CA 94720, USA
}
   \date{Received September 15, 1996; accepted March 16, 1997}

 
  \abstract
{Observations of ultra-hot Jupiters indicate the existence of thermal inversion in their atmospheres with dayside temperatures greater than 2200\,K. Various physical mechanisms such as non-local thermal equilibrium, cloud formation, disequilibrium chemistry, ionisation, hydrodynamic waves and associated energy, have been omitted in previous spectral retrievals while they play an important role on the thermal structure of their upper atmospheres.
}
{We aim at exploring the atmospheric properties of WASP-19b to understand its largely featureless thermal spectra using a state-of-the-art atmosphere code that includes a detailed treatment of the most important physical and chemical processes at play in such atmospheres. 
}
{We used the one-dimensional line-by-line radiative transfer code {\tt PHOENIX} in its spherical symmetry configuration including the BT-Settl cloud model and C/O disequilibrium chemistry to analyse the observed thermal spectrum of WASP-19b. 
}
{We find evidence for a thermal inversion in the dayside atmosphere of the highly irradiated ultra-hot Jupiter WASP-19b with $\teq$ $\sim$ 2700\,K. At these high temperatures we find that H$_2$O dissociates thermally at pressure below $10^{-2}$ bar. The inverted temperature-pressure profiles of WASP-19b show the evidence of CO emission features at 4.5 $\mu$m in its secondary eclipse spectra.
}
{We find that the atmosphere of WASP-19b is thermally inverted. We infer that the thermal inversion is due to the strong impinging radiation. We show that H$_2$O is partially dissociated in the upper atmosphere above about $\tau$ = $10^{-2}$, but is still a significant contributor to the infrared-opacity, dominated by CO. The high-temperature and low-density conditions cause H$_2$O to have a flatter opacity profile than in non-irradiated brown dwarfs. Altogether these factors makes H$_2$O more difficult to identify in WASP-19b. We suggest that state-of-the-art {\tt PHOENIX} model atmosphere code is well-suited to the study of this new class of extrasolar planets, that is the ultra-hot Jupiters.
}


   \maketitle
%

\section{Introduction}

More than 4000 exoplanets have been discovered so far via detection techniques such as the transit and radial velocity methods. A large number of them falls into the category of hot Jupiters and are normally found around around A, F and G type stars \citep{Zellem2017}.  In recent times, a new class of ultra-hot Jupiters has also been identified. Due to their close proximity to their host star, these ultra-hot Jupiters show distinct characteristics  compared with hot Jupiters, such as a higher rate of transit and eclipse. Ultra-hot Jupiters with an orbital separation of less than 0.05 AU face the highest level of irradiation from their host stars. With large atmospheric scale heights they display dayside temperatures $\geq$\,2200 K. Their short orbital periods and high-temperature result in a high planet-to-star contrast ratio, particularly in the mid-infrared. This makes them ideal candidates to probe their atmospheres \citep{Burrows2004}. 

The atmospheres of hot Jupiters are predominantly composed of molecular hydrogen and atomic helium. Other significant molecules possibly found in their atmospheres are CO$_2$, C$_2$H$_2$, and HCN \citep{Madhusudhan2012}. Recent study by \cite{Lothringer2018} showed that various atomic and molecular species such as O, C, N, Fe, Mg, H$_2$O, CO, CH$_4$, N$_2$, NH$_3$ including SiO and metal hydrides are also present at high temperature in the atmospheres of hot Jupiters.

Wavelength dependence of the transit radii observations provides a measure of a planet's atmospheric composition \citep{Fortney2003}. For example, atomic sodium is detected from the transit radii observations of hot-Jupiter HD209458b \citep{Charbonneau2002} whereas H$_2$O has been detected in the atmosphere of both HD189733b \citep{Tinetti2007} and HD209458b \citep{Barman2007}. Observations of ultra-hot Jupiters have also revealed information about the thermal structure of their atmospheres \citep{Madhusudhan2014}.

Depending on the presence of inverted or non-inverted thermal structure profile of their atmospheres, ultra-hot Jupiters can display emission or absorption features in their spectra. They may also show a blackbody spectrum in presence of an isothermal profile \citep{Fortney2008, Line2016}. The nature of the absorbers responsible for the thermal inversion in their atmospheres is not yet known, though the presence of vanadium oxide (VO) and titanium oxide (TiO) has been proposed to be cause for this \citep{Burrows2008, Fortney2008}. The observed absorption broadband features in their optical transmission spectrum between 0.62 and 0.8 $\mu$m can possibly be explained by these absorbers \citep{Desert2008}. Processes such as vertical mixing that prevent the temperature inversion from occurring in their atmosphere are described by \cite{Spiegel2009} and \cite{Parmentier2013, Parmentier2016}, whereas the role of stellar activity for the thermal inversion in the atmosphere of hot-Jupiters is studied by \cite{Knutson2010}. 

Evidences for thermal inversion and diverse characteristics of the atmospheres of ultra-hot Jupiters are also imprinted in their near-infrared (NIR) observations. For example, TiO has been detected as an emission feature in WASP-33b \citep{Haynes2015, Nugroho2017} and as an absorption feature in WASP-19b \citep{Sedaghati2017}. Emission features due to H$_2$O and VO have also been detected in WASP-121b \citep{Evans2017}. The presence of thermal inversion without H$_2$O, TiO, and VO has been noticed in the low resolution secondary eclipse spectra of WASP-18b \citep{Sheppard2017, Arcangeli2018}. Their study provides evidence for CO emission at 4.5 $\mu$m, implying an inverted thermal structure. On the other hand, no clear evidence for emission or absorption features of H$_2$O is found in HAT-P-7b \citep{Mansfield2018}, WASP-12b \citep{Swain2013}, and WASP-103b \citep{Kreidberg2018}. These planets show instead a thermal blackbody spectra. A handful of extremely hot-Jupiters also do not show any hint for thermal inversion such as HD209458b, KELT-1b, and Kepler-13Ab \citep{Diamond-Lowe2014, Beatty2017a, Beatty2017b}.

Retrieval techniques are commonly used to constrain the chemical composition, thermal structure, and to explain the featureless spectra of ultra-hot Jupiters \citep{Madhusudhan2009, Line2013,Stevenson2014}. Such methods allowed to detect the presence of CO with high C/O ratio in WASP-18b \citep{Sheppard2017} and to derive a sub-solar metallicity associated with an oxygen-rich composition in WASP-33b \citep{Haynes2015}. The forward modelling approach by \cite{Arcangeli2018},  \cite{Lothringer2018}, and  \cite{Parmentier2018} explains the lack of strong H$_2$O spectral features in the ultra-hot Jupiters due to molecular dissociation. Considering H-- opacity and the thermal dissociation of H$_2$O, \cite{Arcangeli2018}, \cite{Kreidberg2018}, \cite{Kitzmann2018}, and \cite{Mansfield2018} explained the featureless spectra in some ultra-hot Jupiters for example HAT-P-7b, KELT-9b, WASP-12b, WASP-121b, and WASP-18b. Although the  retrieval techniques are useful to quantify the atmospheric abundances of ultra-hot Jupiters, the results demonstrate a large diversity in their chemical composition and thermal structures, which could lead to biased conclusions. 

In this paper, we investigate the atmospheric properties of the ultra-hot Jupiter WASP-19b. WASP-19b is among the objects that exhibit weaker spectral features compared to their cooler counterparts. WASP-19b has a shorter orbital period and, thus, receives extreme irradiation from its host star. Its atmosphere is then of particular interest to study. We argue that proper chemistry and opacity  play a key role in understanding its extreme hot atmosphere without invoking unusual abundances.

\section{Model Construction}

To model the atmosphere of WASP-19b, we have used the BT-Settl model atmosphere described in \cite{Allard2012} and  \cite{Rajpurohit2012a}. BT-Settl is a state-of-the-art model atmosphere code which has been extremely successful to reproduce the properties of stellar to sub-stellar objects including cool brown dwarfs \citep{Rajpurohit2012a, Rajpurohit2013}. The current version of the BT-Settl model uses H$_2$O along with other molecular line lists such as FeH, CrH, TiO, VO, CaH, NH$_3$, Mg, CO$_2$ and CO. Collision induced absorption (CIA) from H$_2$ collisions with H$_2$, He, CH$_4$, N$_2$, and Ar along with CO$_2$-CO$_2$, Ar-CH$_4$, and CH$_4$-CH$_4$ are also included (for more detail see \cite{Allard2012}). The BT-Settl model accounts for many continuous opacity sources, including bound-free opacity from H, H--, He, C, N, O, Na, Mg, Al, Si, S, Ca, and Fe, free-free opacity from H, Mg, and Si, and scattering from e-- , H, He, and H$_2$. The model includes the opacities of more than 100 molecular species, including many molecules, isotopes, atomic species and their ionised states. Detailed profiles for the alkali lines as described in \cite{Unsold1968}, \cite{Valenti1996}, and \cite{Allard2007} and  approximation is utilised for the atomic damping constants with a correction factor to the widths of 2.5 for the non-hydrogenic atoms. Several important atomic transitions, such as the alkali, Ca I, and Ca II resonance lines along with more accurate broadening data for neutral hydrogen collisions by \cite{Barklem2000} have been included. 

The BT-Settl model also accounts for disequilibrium chemistry for C/O and N  \citep{Graven1954, Visscher2010, Allard2012}. The reference solar elemental abundances are derived from \cite{Caffau2011}. These models are computed with the version 15.5 of {\tt PHOENIX}, a multi-purpose atmosphere code  \citep{Allard2001}. {\tt PHOENIX} solves the radiative transfer in 1D spherical symmetry with irradiation such that flux is conserved at each layer. The basic assumption in {\tt PHOENIX} while solving radiative transfer is hydrostatic equilibrium with convection using the mixing length theory, and a sampling treatment of the opacities \citep[see][]{Allard2013}.

The atmosphere model for WASP-19b irradiated by the host star of spectral type G5 with an orbital separation of 0.0163 AU has been computed by considering boundary conditions as described in \cite{Barman2001}. A pre-computed converged non-irradiated model structure at 400 K is chosen from \cite{Allard2012} as the initial model structure to start with. The temperature is decreased down to 100 K to achieve the model structure. A temperature  change of less than 1 K at every depth point in the thermal structure is considered as the criterion for the converged model. Effective temperature ($\teff$) of 5500 K, surface gravity ($\logg$) of 4.5, and [M/H]) = 0 have been taken as input parameters for the host star.
  
 \begin{figure}[!htbp]
 \centering
    \includegraphics[width=0.50\textwidth]{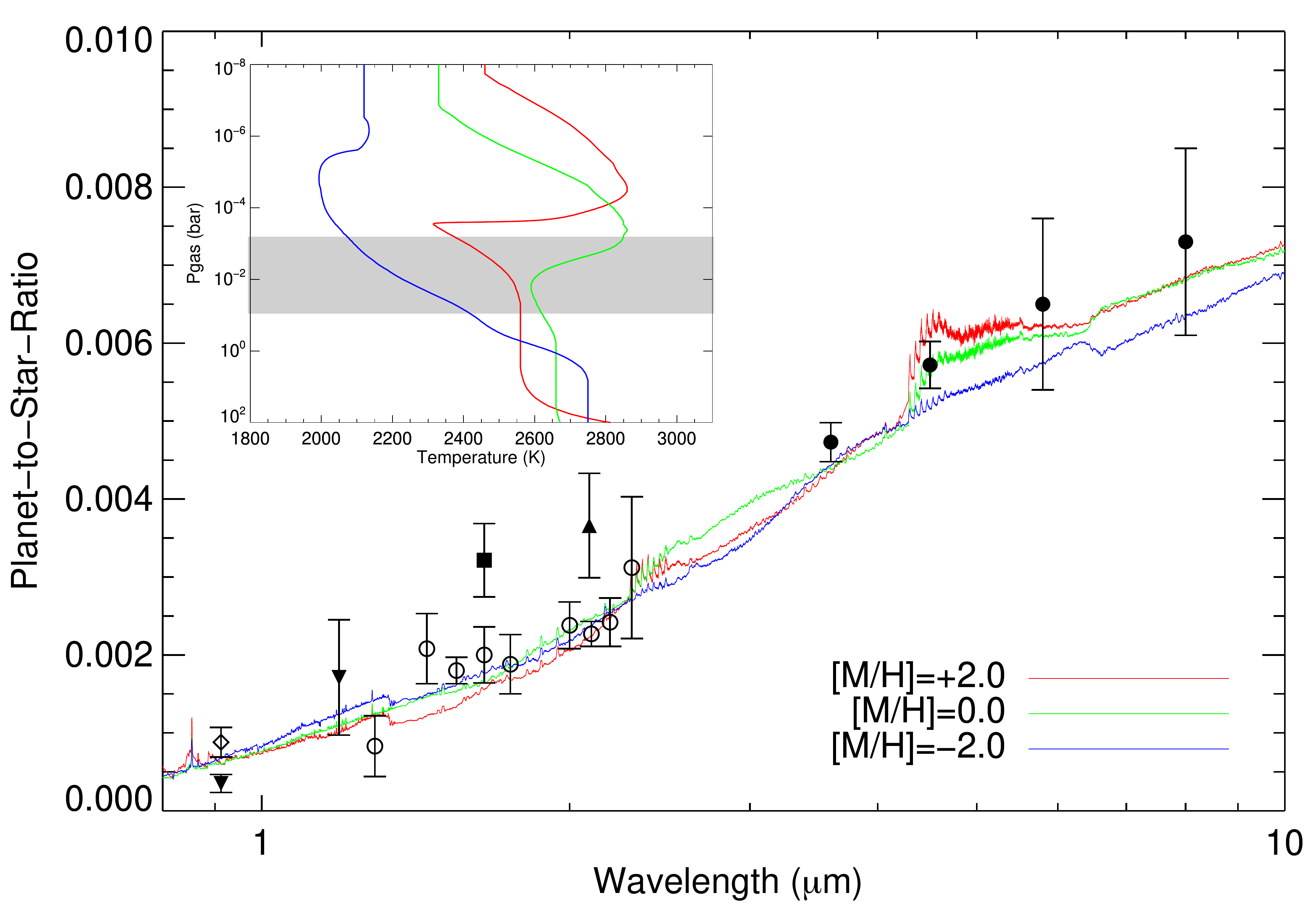}
  \vspace{-0.3cm}
   \caption{Comparison of observed thermal spectra of WASP-19b with the modelled spectra calculated using {\tt PHOENIX} (solid line) at different metallicity with an average dayside heat redistribution ($f$ = 0.5).  The corresponding temperature-pressure profile is shown in the inset image (the grey shaded region corresponds to the structure inversion). The observed thermal spectrum of WASP-19b presented in this paper is adopted from \cite{Anderson2010, Anderson2013} (filled square and filled circles), \cite{Gibson2010} (filled upward triangle), \cite{Bean2013} (open circles), \cite{Burton2012} (open diamond), and \cite{Lendl2013} (filled downward triangles). The emission features at 4.5 $\mu$m are due to CO in presence of thermal inversion.}
\label{fig1}
\end{figure}

To explore and investigate the atmospheric properties of WASP-19b, models with uniform heat redistribution with $f$ = 0.5 across the entire dayside have been calculated with slanted rays for metallicity at --2.0, 0.0, and  +2.0 dex. Here the factor $f$ measures the uniform heat redistribution where $f$ = 0.5 indicates heat redistribution across the entire dayside while $f$ = 1 indicates full heat redistribution. The model is calculated on an optical depth grid of 250 layers in log-space for $\tau$ = $10^{-9}$ to $10^{3}$ at the peak of the spectral energy distribution which corresponds to pressures of $10^{-12}$ to $\sim$10 bars. In the computation of the atmospheric model, some non-local thermal equilibrium (NLTE) processes have also been included for a small set of elements and level numbers (H I, He I, Li I, C I, Ni I, O I, Ne I, Na I, Na II, Mg I, Mg II, Si I, S I, K I, K II, Ca I, Ca II, Rb I), which affect the entire atmosphere. {\tt PHOENIX} has been used to calculate both the planetary and stellar spectra from 10 to $10^{6}$ \AA. The planetary, star and orbital parameters used for the computation irradiated model of WASP-19b are  summarised in Table~\ref{tab}. 

\begin{table}[!htbp]
  \center
  \caption{\label{tab} Planetary, stellar and orbital parameters used in the model are from \cite{Butler2006}, \cite{Southworth2007}, \cite{Hebb2010} and \cite{Hellier2011}.} 
  \begin{tabular}{cc}
    \hline
    \hline
    \noalign{\smallskip}
    Parameter & Value \\
    \noalign{\smallskip}
    \hline
    \noalign{\smallskip}
    $a$ & 0.0163 AU \\[1mm]
     \noalign{\smallskip}
    R\textsubscript{[p]}  			& 1.386 ({R\textsubscript{Jupiter}}) \\[1mm]
     \noalign{\smallskip}
    $\logg$\textsubscript{[p]} 		& 3.17 [cm/sec$^2$]\\[1mm]
     \noalign{\smallskip}
    $\tint$\textsubscript{[p]}   		& 100 [K]\\[1mm]
     \noalign{\smallskip}
     $\teff$\textsubscript{[$\star$]}   	& 5500 [K]\\[1mm]
     \noalign{\smallskip}
    $\logg$\textsubscript{[$\star$]}		& 4.5 [cm/sec$^2$]\\[1mm]
     \noalign{\smallskip}
     R\textsubscript{[$\star$]}   		& 0.990 ({R\textsubscript{$\sun$}}) \\[1mm]
     \noalign{\smallskip}
      [M/H]\textsubscript{[$\star$]}   	& 0.0 \\[1mm]
     \hline
  \end{tabular}
\end{table}

\section{Results and Discussion}

A thermal spectrum of WASP-19b, from \cite{Anderson2010}, \cite{Gibson2010}, \cite{Burton2012}, \cite{Anderson2013}, \cite{Bean2013}, and \cite{Lendl2013}, is shown in Fig~\ref{fig1}.  The planet-to-star flux ratio at 4.5 $\mu$m is higher than the one at 3.6 $\mu$m {\tt Spitzer}/IRAC channel. Moreover, at higher wavelengths, namely at 5.8 and 8.0 $\mu$m, the planet-to-star flux ratio is even higher than 4.5 $\mu$m. This clearly indicates the presence of excess emission in some photometric bands over others. 

The atmosphere of WASP-19b has been modelled with the parameters given in Table~\ref{tab} using {\tt PHOENIX}. It shows the presence of thermal inversion (see inset diagram, Fig.~\ref{fig1}) and the equilibrium temperature is found to be $\sim$ 2700 K. The comparison between observational data and our theoretical modelling (Fig~\ref{fig1}) indicates that the excess planet-to-star flux ratio at 4.5 $\mu$m is due to the presence of CO, which in turn is caused by thermal inversion that exists in the atmosphere. The corresponding pressure range, as probed by the secondary eclipse, is $10^{-2}$ to $10^{-3}$ bar. Previous studies on WASP-19b did not show any evidence of thermal inversion in its atmosphere \citep{Anderson2013, Bean2013}. However, these studies were mainly based on retrieval techniques where abundances and thermal structure were fitted to the data.


 \begin{figure*}[!htbp]
 \centering
 \includegraphics[height=6cm,width=17cm]{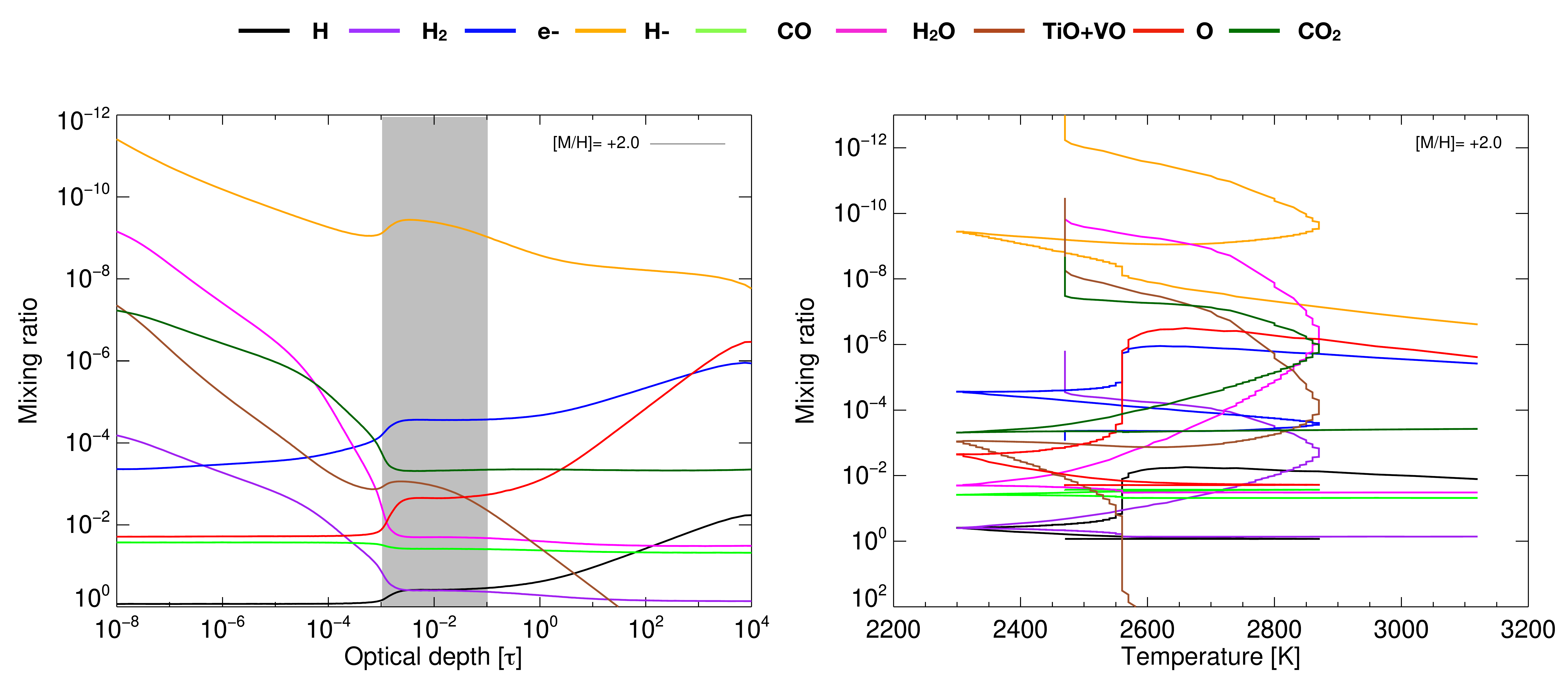}
  \includegraphics[height=6cm,width=17cm]{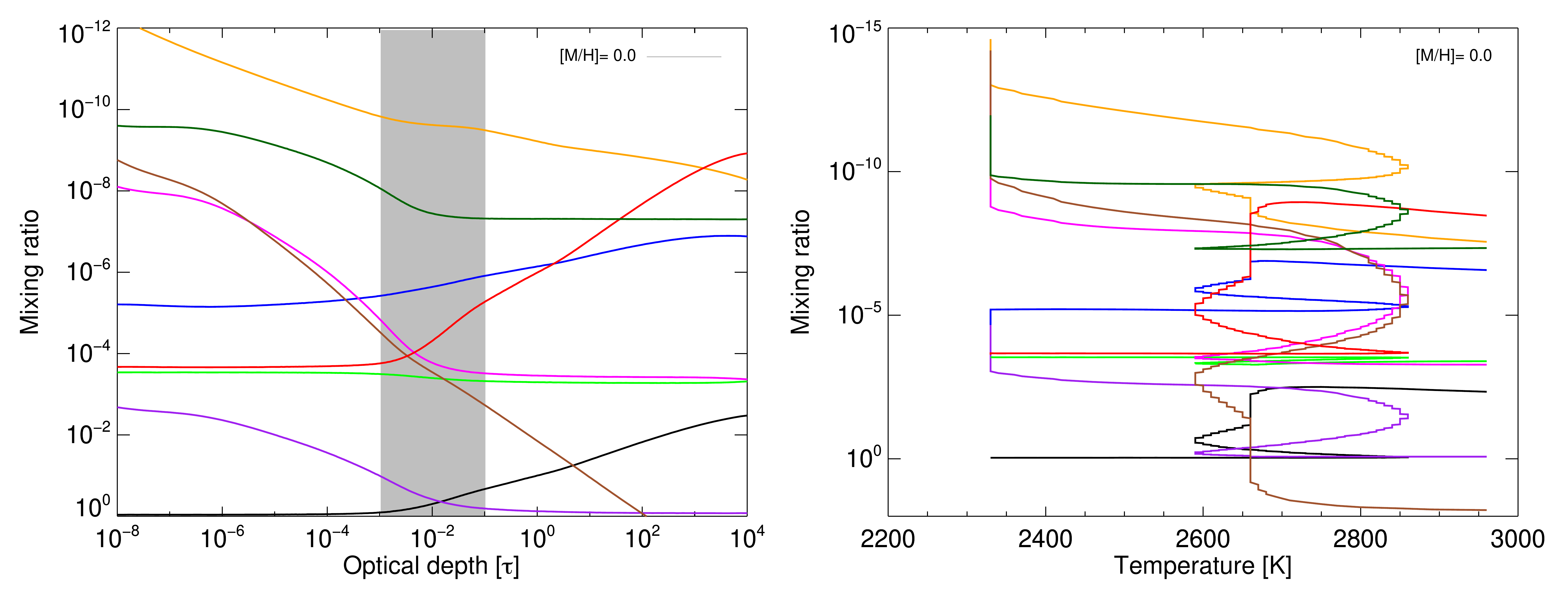}
   \includegraphics[height=6cm,width=17cm]{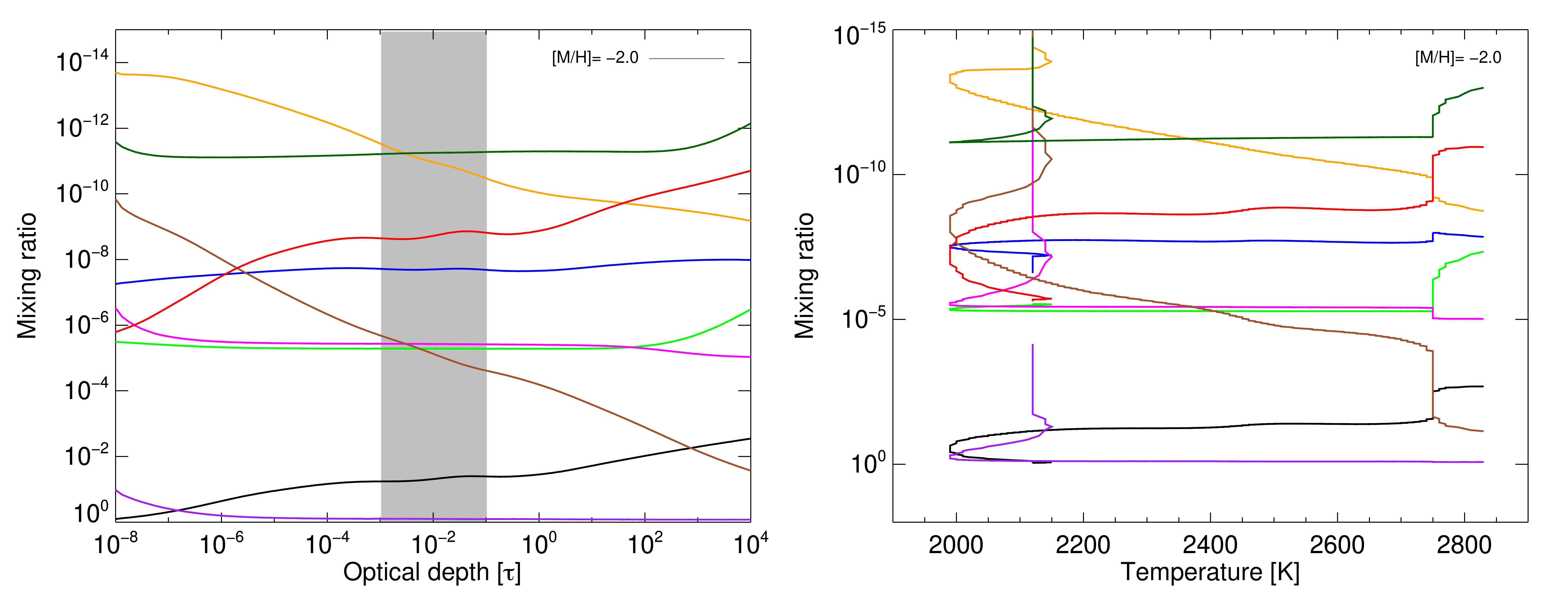}
  \vspace{-0.2cm}
   \caption{Concentration of important absorbing molecules and neutral atoms as a function of optical depth and temperature at [M/H] = +2.0 (top), [M/H] = 0.0 (middle), and [M/H] = -2.0 (bottom). These species deplete at the given pressure range for the irradiated hot Jupiter WASP-19b (see main text). The grey shaded region (left panels) corresponds to the structure inversion and shows the window of the observable atmosphere from which most of the spectrum emerges (in the CO, H$_2$O, TiO molecular bands pseudo continuum). We notice that in the upper atmosphere ($\tau$ $<$  $10^{-2}$), H$_2$O, TiO and VO dissociate whereas neutral atomic oxygen remains constant. We find that CO remain constant throughout the atmosphere of WASP-19b. The temperature inversion (right panels) causes the wiggles along the structure causing CO bands to emerge in emission.}.
 \label{fig2}
\end{figure*}

We investigate the effect of metallicity on the atmosphere of WASP-19b by changing all the heavy-elemental abundances by a factor of two. As shown by the inset image of Fig.\,\ref{fig1}, the thermal inversion becomes stronger as the metallicity increases. This is mainly due to the temperature dependence of the thermal dissociation of various molecules, which changes the photospheric abundances ratio of TiO and H$_2$O at different metallicities \citep{Parmentier2018}. At higher metallicities, the increase in the abundances of various absorbers such as TiO warms the upper atmosphere. This causes the thermal inversion to be even stronger as compared to solar metallicity which leads to strong emission features of CO and shallower features of H$_2$O in their thermal spectra. We find that models with sub-solar metallicities show  signature of a very weak or no thermal inversion  (see Fig.\,\ref{fig1}). This is a result of the decrease of TiO and VO abundances. In most of the hot Jupiters or massive planets, the metallicity is believed to approach the metallicity of the host star \citep{Torres2012, Arcangeli2018}. We also investigate this by comparing the observed thermal spectra of the dayside atmosphere of WASP-19b with the models at [M/H] = -2.0, 0.0 and +2.0. Our results show that thermal spectra of WASP-19b can be explained with a model at solar metallicity with thermal inversion without invoking unusual abundances.

In case of ultra-hot Jupiters, the molecular species required to explain the featureless observations are physically plausible.  At extremely high temperature, the various molecules responsible for the radiative cooling in their atmosphere do not exist. In Fig.\,\ref{fig2}, we show the mixing ratio of the most important atomic and molecular species in the atmosphere of WASP-19b at different metallicity as a function of optical depth and temperature. We see that H$_2$ dissociates in the upper atmosphere due to impinging radiation from about $\tau$ = $10^{-2}$, but not enough to prevent the molecular hydrogen atmosphere. Due to the impinging radiation, free electrons are available and constant over most of the upper photosphere. Neutral atomic oxygen is fully locked to the very stable CO molecule in the deep atmosphere ($P$ $>$ $10^{-2}$ bar), but becomes as abundant as CO in the upper atmosphere. This is the result of partial dissociation of CO, and other oxygen--bearing molecules. While CO is quasi-constant throughout the atmosphere. We find that H$_2$ and CO are the most abundant molecules, while CO, TiO and VO are the most important molecular opacity sources in the atmosphere of WASP-19b. 

As shown Fig.\,\ref{fig2} (left panels), we find that the variation of atomic and molecular abundances, together with the strong impinging radiation, contribute to the thermal inversion in the atmosphere of WASP-19b. At 0.0163 AU, the thermal structure of WASP-19b is too hot for the formation of the BT-Settl clouds, whereas the presence of CO/CO$_2$ reveals the disequilibrium chemistry. We find the formation of a limited amount of CO$_2$ in disequilibrium chemistry, but not abundant enough to participate significantly in the CO and H$_2$O balance. We also show that TiO and alkali doublets, seen in early to mid-type brown dwarfs, are the main opacities in the optical to near-IR spectrum of WASP-19b. It is evident from Fig.\,\ref{fig2} (right panels) that the temperature inversion causes the wiggles in the concentrations along the structure causing CO bands to appear in emission. 

\begin{figure}[!htbp]
 \centering
    \includegraphics[width=0.50\textwidth]{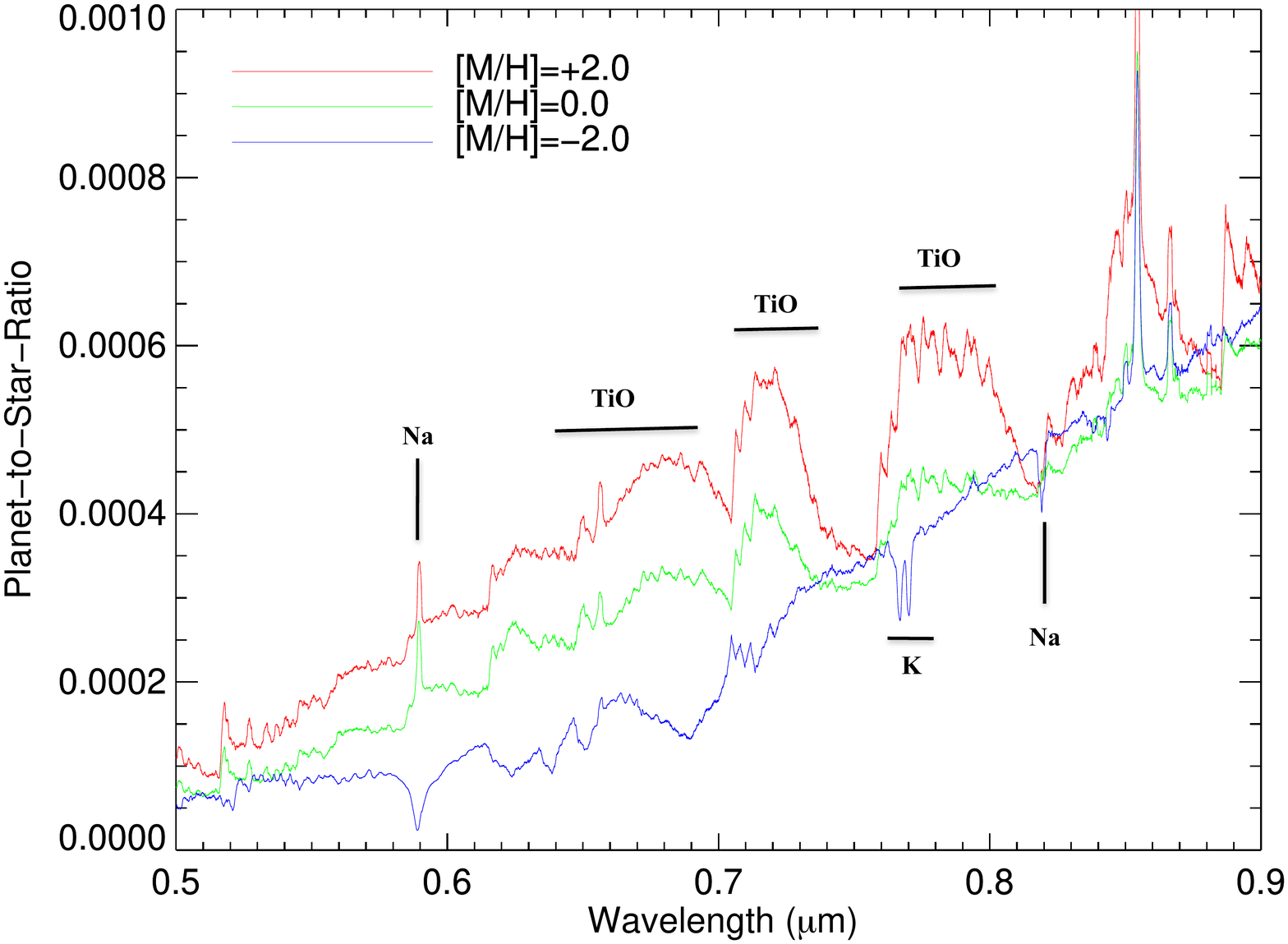}
  \vspace{-0.7cm}
   \caption{Synthetic spectra of WASP-19b in the optical range calculated using {\tt PHOENIX} at different metallicities with an average dayside heat redistribution (f = 0.5). Labels of the most prominent atomic and molecular features remaining important opacity sources, even at low concentrations, are indicated.
}
\label{fig3}
\end{figure}

Figure\,\ref{fig3} shows the synthetic spectra of WASP-19b in the optical at different metallicities. We see that despite extremely low abundances, similar to brown dwarfs and irradiated hot-Jupiter atmospheres \citep{Allard2001, Barman2001, Barman2002}, the TiO cross-sections are strong enough to preserve TiO as the main optical to near-IR opacity, along with alkali atomic fundamental transitions. These are the most important opacity sources, together with CO in the infrared, that survive the immense heat impinging of the planet atmosphere. The thermal inversion potentially hides pseudo-continuum opacities such as H$_2$O, and the high temperatures do not allow triatomic to remain stable. Also at given resolution and at such high temperature conditions, H$_2$O has a much flatter opacity profile making it more difficult to recognise, especially at those spectral resolutions.


\section{Conclusions}

We have used the state-of-the-art 1D NLTE opacity sampling model atmosphere code {\tt PHOENIX} to study the atmosphere of WASP-19b. This model has been successfully used to study the atmosphere of cool stars, brown dwarfs and extrasolar planets. The temperature-pressure profile of WASP-19b computed using {\tt PHOENIX} shows the presence of thermal inversion. Our model computed at solar metallicity successfully reproduces the observed photometry of WASP-19b without the need for non-solar composition. The secondary eclipse of WASP-19b shows evidence for CO emission features at 4.5 $\mu$m, but however no sign of H$_2$O. We find that these features are the results of thermal dissociation and thermal inversion due to the strong impinging radiation. Also, the H$_2$O pseudo-continuum is much smoother at the concerned temperatures and densities, making it more difficult to identify at those extremely coarse spectral resolutions. Our results further strengthen the fact that the family of ultra-hot Jupiters commonly exhibit thermal inversions.   

At longer wavelengths (above 1.4 \,$\mu$m) and at extremely hot temperatures (2200--2800 K), a significant amount of H$_2$O gets thermally dissociated at a pressure below $10^{-2}$ bar. At a temperature above 2200 K, TiO and VO do remain important opacity sources, even at low concentrations. At longer wavelengths, CO is the only molecule with its strong triple bond to be abundant below $10^{-2}$ bar. This molecule does not dissociate and its emission features appear at 4.5 $\mu$m. We suggest that the actual reason for the drop in H$_2$O in ultra-hot Jupiters is the partial thermal dissociation of this molecule and the resulting thermal inversion which shapes the thermal spectra of ultra-hot Jupiters. This makes H$_2$O a poor diagnosis of the C/O ratio. Also at such high dayside temperature ($>$ 2200 K), and at that resolution, the opacity profile of H$_2$O is spectrally much flatter and difficult to recognise.

\begin{acknowledgements}
We thank the anonymous referee for providing comments and suggestions which helped improve the clarity and conciseness of the paper. AR are especially grateful to Mudit Kumar Srivastava from Physical Research Laboratory (PRL) for providing feedback on the manuscript. The computations were performed on the HPC resources at the Physical Research Laboratory (PRL). The research leading to these results has received funding from the French "Programme National de Physique Stellaire" and the Programme National de Planetologie of CNRS (INSU). The computations were performed at the {\sl P\^ole Scientifique de Mod\'elisation Num\'erique} (PSMN) at the {\sl \'Ecole Normale Sup\'erieure} (ENS) in Lyon, and at the {\sl Gesellschaft f{\"u}r Wissenschaftliche Datenverarbeitung G{\"o}ttingen} in collaboration with the Institut f{\"u}r Astrophysik G{\"o}ttingen.  O.M. acknowledges support from CNES. 
\end{acknowledgements}

\bibliographystyle{aa}
\bibliography{exoplanet_ref}

\begin{thebibliography}{59}
\expandafter\ifx\csname natexlab\endcsname\relax\def\natexlab#1{#1}\fi

\bibitem[{{Allard} {et~al.}(2007){Allard}, {Allard}, {Homeier}, {Kielkopf},
  {McCaughrean}, \& {Spiegelman}}]{Allard2007}
{Allard}, F., {Allard}, N.~F., {Homeier}, D., {et~al.} 2007, \aap, 474, L21

\bibitem[{{Allard} {et~al.}(2001){Allard}, {Hauschildt}, {Alexander},
  {Tamanai}, \& {Schweitzer}}]{Allard2001}
{Allard}, F., {Hauschildt}, P.~H., {Alexander}, D.~R., {Tamanai}, A., \&
  {Schweitzer}, A. 2001, \apj, 556, 357

\bibitem[{{Allard} {et~al.}(2013){Allard}, {Homeier}, {Freytag},
  {Schaffenberger}, {}, \& {Rajpurohit}}]{Allard2013}
{Allard}, F., {Homeier}, D., {Freytag}, B., {et~al.} 2013, Memorie della
  Societa Astronomica Italiana Supplementi, 24, 128

\bibitem[{{Allard} {et~al.}(2012){Allard}, {Homeier}, {Freytag}, \&
  {Sharp}}]{Allard2012}
{Allard}, F., {Homeier}, D., {Freytag}, B., \& {Sharp}, C.~M. 2012, in EAS
  Publications Series, Vol.~57, EAS Publications Series, ed. C.~{Reyl{\'e}},
  C.~{Charbonnel}, \& M.~{Schultheis}, 3--43

\bibitem[{{Anderson} {et~al.}(2010){Anderson}, {Gillon}, {Maxted}, {Barman},
  {Collier Cameron}, {Hellier}, {Queloz}, {Smalley}, \&
  {Triaud}}]{Anderson2010}
{Anderson}, D.~R., {Gillon}, M., {Maxted}, P.~F.~L., {et~al.} 2010, \aap, 513,
  L3

\bibitem[{{Anderson} {et~al.}(2013){Anderson}, {Smith}, {Madhusudhan},
  {Wheatley}, {Collier Cameron}, {Hellier}, {Campo}, {Gillon}, {Harrington},
  {Maxted}, {Pollacco}, {Queloz}, {Smalley}, {Triaud}, \&
  {West}}]{Anderson2013}
{Anderson}, D.~R., {Smith}, A.~M.~S., {Madhusudhan}, N., {et~al.} 2013, \mnras,
  430, 3422

\bibitem[{{Arcangeli} {et~al.}(2018){Arcangeli}, {D{\'e}sert}, {Line}, {Bean},
  {Parmentier}, {Stevenson}, {Kreidberg}, {Fortney}, {Mansfield}, \&
  {Showman}}]{Arcangeli2018}
{Arcangeli}, J., {D{\'e}sert}, J.-M., {Line}, M.~R., {et~al.} 2018, \apjl, 855,
  L30

\bibitem[{{Barklem} {et~al.}(2000){Barklem}, {Piskunov}, \&
  {O'Mara}}]{Barklem2000}
{Barklem}, P.~S., {Piskunov}, N., \& {O'Mara}, B.~J. 2000, \aap, 363, 1091

\bibitem[{{Barman}(2007)}]{Barman2007}
{Barman}, T. 2007, \apjl, 661, L191

\bibitem[{{Barman} {et~al.}(2001){Barman}, {Hauschildt}, \&
  {Allard}}]{Barman2001}
{Barman}, T.~S., {Hauschildt}, P.~H., \& {Allard}, F. 2001, \apj, 556, 885

\bibitem[{{Barman} {et~al.}(2002){Barman}, {Hauschildt}, {Schweitzer},
  {Stancil}, {Baron}, \& {Allard}}]{Barman2002}
{Barman}, T.~S., {Hauschildt}, P.~H., {Schweitzer}, A., {et~al.} 2002, \apjl,
  569, L51

\bibitem[{{Bean} {et~al.}(2013){Bean}, {D{\'e}sert}, {Seifahrt}, {Madhusudhan},
  {Chilingarian}, {Homeier}, \& {Szentgyorgyi}}]{Bean2013}
{Bean}, J.~L., {D{\'e}sert}, J.-M., {Seifahrt}, A., {et~al.} 2013, \apj, 771,
  108

\bibitem[{{Beatty} {et~al.}(2017{\natexlab{a}}){Beatty}, {Madhusudhan},
  {Pogge}, {Chung}, {Bierlya}, {Gaudi}, \& {Latham}}]{Beatty2017a}
{Beatty}, T.~G., {Madhusudhan}, N., {Pogge}, R., {et~al.} 2017{\natexlab{a}},
  \aj, 154, 242

\bibitem[{{Beatty} {et~al.}(2017{\natexlab{b}}){Beatty}, {Madhusudhan},
  {Tsiaras}, {Zhao}, {Gilliland}, {Knutson}, {Shporer}, \&
  {Wright}}]{Beatty2017b}
{Beatty}, T.~G., {Madhusudhan}, N., {Tsiaras}, A., {et~al.} 2017{\natexlab{b}},
  \aj, 154, 158

\bibitem[{{Burrows} {et~al.}(2008){Burrows}, {Budaj}, \&
  {Hubeny}}]{Burrows2008}
{Burrows}, A., {Budaj}, J., \& {Hubeny}, I. 2008, \apj, 678, 1436

\bibitem[{{Burrows} {et~al.}(2004){Burrows}, {Sudarsky}, \&
  {Hubeny}}]{Burrows2004}
{Burrows}, A., {Sudarsky}, D., \& {Hubeny}, I. 2004, \apj, 609, 407

\bibitem[{{Burton} {et~al.}(2012){Burton}, {Watson}, {Littlefair}, {Dhillon},
  {Gibson}, {Marsh}, \& {Pollacco}}]{Burton2012}
{Burton}, J.~R., {Watson}, C.~A., {Littlefair}, S.~P., {et~al.} 2012, \apjs,
  201, 36

\bibitem[{{Butler} {et~al.}(2006){Butler}, {Wright}, {Marcy}, {Fischer},
  {Vogt}, {Tinney}, {Jones}, {Carter}, {Johnson}, {McCarthy}, \&
  {Penny}}]{Butler2006}
{Butler}, R.~P., {Wright}, J.~T., {Marcy}, G.~W., {et~al.} 2006, \apj, 646, 505

\bibitem[{{Caffau} {et~al.}(2011){Caffau}, {Ludwig}, {Steffen}, {Freytag}, \&
  {Bonifacio}}]{Caffau2011}
{Caffau}, E., {Ludwig}, H.-G., {Steffen}, M., {Freytag}, B., \& {Bonifacio}, P.
  2011, \solphys, 268, 255

\bibitem[{{Charbonneau} {et~al.}(2002){Charbonneau}, {Brown}, {Noyes}, \&
  {Gilliland}}]{Charbonneau2002}
{Charbonneau}, D., {Brown}, T.~M., {Noyes}, R.~W., \& {Gilliland}, R.~L. 2002,
  \apj, 568, 377

\bibitem[{{D{\'e}sert} {et~al.}(2008){D{\'e}sert}, {Vidal-Madjar}, {Lecavelier
  Des Etangs}, {Sing}, {Ehrenreich}, {H{\'e}brard}, \& {Ferlet}}]{Desert2008}
{D{\'e}sert}, J.~M., {Vidal-Madjar}, A., {Lecavelier Des Etangs}, A., {et~al.}
  2008, \aap, 492, 585

\bibitem[{{Diamond-Lowe} {et~al.}(2014){Diamond-Lowe}, {Stevenson}, {Bean},
  {Line}, \& {Fortney}}]{Diamond-Lowe2014}
{Diamond-Lowe}, H., {Stevenson}, K.~B., {Bean}, J.~L., {Line}, M.~R., \&
  {Fortney}, J.~J. 2014, \apj, 796, 66

\bibitem[{{Evans} {et~al.}(2017){Evans}, {Sing}, {Kataria}, {Goyal}, {Nikolov},
  {Wakeford}, {Deming}, {Marley}, {Amundsen}, {Ballester}, {Barstow},
  {Ben-Jaffel}, {Bourrier}, {Buchhave}, {Cohen}, {Ehrenreich}, {Garc{\'\i}a
  Mu{\~n}oz}, {Henry}, {Knutson}, {Lavvas}, {Lecavelier Des Etangs}, {Lewis},
  {L{\'o}pez-Morales}, {Mandell}, {Sanz-Forcada}, {Tremblin}, \&
  {Lupu}}]{Evans2017}
{Evans}, T.~M., {Sing}, D.~K., {Kataria}, T., {et~al.} 2017, \nat, 548, 58

\bibitem[{{Fortney} {et~al.}(2008){Fortney}, {Lodders}, {Marley}, \&
  {Freedman}}]{Fortney2008}
{Fortney}, J.~J., {Lodders}, K., {Marley}, M.~S., \& {Freedman}, R.~S. 2008,
  \apj, 678, 1419

\bibitem[{{Fortney} {et~al.}(2003){Fortney}, {Sudarsky}, {Hubeny}, {Cooper},
  {Hubbard}, {Burrows}, \& {Lunine}}]{Fortney2003}
{Fortney}, J.~J., {Sudarsky}, D., {Hubeny}, I., {et~al.} 2003, \apj, 589, 615

\bibitem[{{Gibson} {et~al.}(2010){Gibson}, {Aigrain}, {Pollacco}, {Barros},
  {Hebb}, {Hrudkov{\'a}}, {Simpson}, {Skillen}, \& {West}}]{Gibson2010}
{Gibson}, N.~P., {Aigrain}, S., {Pollacco}, D.~L., {et~al.} 2010, \mnras, 404,
  L114

\bibitem[{Graven \& Long(1954)}]{Graven1954}
Graven, W.~M. \& Long, F.~J. 1954, Journal of the American Chemical Society,
  76, 2602

\bibitem[{{Haynes} {et~al.}(2015){Haynes}, {Mandell}, {Madhusudhan}, {Deming},
  \& {Knutson}}]{Haynes2015}
{Haynes}, K., {Mandell}, A.~M., {Madhusudhan}, N., {Deming}, D., \& {Knutson},
  H. 2015, \apj, 806, 146

\bibitem[{{Hebb} {et~al.}(2010){Hebb}, {Collier-Cameron}, {Triaud}, {Lister},
  {Smalley}, {Maxted}, {Hellier}, {Anderson}, {Pollacco}, {Gillon}, {Queloz},
  {West}, {Bentley}, {Enoch}, {Haswell}, {Horne}, {Mayor}, {Pepe}, {Segransan},
  {Skillen}, {Udry}, \& {Wheatley}}]{Hebb2010}
{Hebb}, L., {Collier-Cameron}, A., {Triaud}, A.~H.~M.~J., {et~al.} 2010, \apj,
  708, 224

\bibitem[{{Hellier} {et~al.}(2011){Hellier}, {Anderson}, {Collier-Cameron},
  {Miller}, {Queloz}, {Smalley}, {Southworth}, \& {Triaud}}]{Hellier2011}
{Hellier}, C., {Anderson}, D.~R., {Collier-Cameron}, A., {et~al.} 2011, \apjl,
  730, L31

\bibitem[{{Kitzmann} {et~al.}(2018){Kitzmann}, {Heng}, {Rimmer}, {Hoeijmakers},
  {Tsai}, {Malik}, {Lendl}, {Deitrick}, \& {Demory}}]{Kitzmann2018}
{Kitzmann}, D., {Heng}, K., {Rimmer}, P.~B., {et~al.} 2018, \apj, 863, 183

\bibitem[{{Knutson} {et~al.}(2010){Knutson}, {Howard}, \&
  {Isaacson}}]{Knutson2010}
{Knutson}, H.~A., {Howard}, A.~W., \& {Isaacson}, H. 2010, \apj, 720, 1569

\bibitem[{{Kreidberg} {et~al.}(2018){Kreidberg}, {Line}, {Parmentier},
  {Stevenson}, {Louden}, {Bonnefoy}, {Faherty}, {Henry}, {Williamson},
  {Stassun}, {Beatty}, {Bean}, {Fortney}, {Showman}, {D{\'e}sert}, \&
  {Arcangeli}}]{Kreidberg2018}
{Kreidberg}, L., {Line}, M.~R., {Parmentier}, V., {et~al.} 2018, \aj, 156, 17

\bibitem[{{Lendl} {et~al.}(2013){Lendl}, {Gillon}, {Queloz}, {Alonso}, {Fumel},
  {Jehin}, \& {Naef}}]{Lendl2013}
{Lendl}, M., {Gillon}, M., {Queloz}, D., {et~al.} 2013, \aap, 552, A2

\bibitem[{{Line} \& {Parmentier}(2016)}]{Line2016}
{Line}, M.~R. \& {Parmentier}, V. 2016, \apj, 820, 78

\bibitem[{{Line} {et~al.}(2013){Line}, {Wolf}, {Zhang}, {Knutson}, {Kammer},
  {Ellison}, {Deroo}, {Crisp}, \& {Yung}}]{Line2013}
{Line}, M.~R., {Wolf}, A.~S., {Zhang}, X., {et~al.} 2013, \apj, 775, 137

\bibitem[{{Lothringer} {et~al.}(2018){Lothringer}, {Barman}, \&
  {Koskinen}}]{Lothringer2018}
{Lothringer}, J.~D., {Barman}, T., \& {Koskinen}, T. 2018, \apj, 866, 27

\bibitem[{{Madhusudhan}(2012)}]{Madhusudhan2012}
{Madhusudhan}, N. 2012, \apj, 758, 36

\bibitem[{{Madhusudhan} {et~al.}(2014){Madhusudhan}, {Knutson}, {Fortney}, \&
  {Barman}}]{Madhusudhan2014}
{Madhusudhan}, N., {Knutson}, H., {Fortney}, J.~J., \& {Barman}, T. 2014, in
  Protostars and Planets VI, ed. H.~{Beuther}, R.~S. {Klessen}, C.~P.
  {Dullemond}, \& T.~{Henning}, 739

\bibitem[{{Madhusudhan} \& {Seager}(2009)}]{Madhusudhan2009}
{Madhusudhan}, N. \& {Seager}, S. 2009, \apj, 707, 24

\bibitem[{{Mansfield} {et~al.}(2018){Mansfield}, {Bean}, {Line}, {Parmentier},
  {Kreidberg}, {D{\'e}sert}, {Fortney}, {Stevenson}, {Arcangeli}, \&
  {Dragomir}}]{Mansfield2018}
{Mansfield}, M., {Bean}, J.~L., {Line}, M.~R., {et~al.} 2018, \aj, 156, 10

\bibitem[{{Nugroho} {et~al.}(2017){Nugroho}, {Kawahara}, {Masuda}, {Hirano},
  {Kotani}, \& {Tajitsu}}]{Nugroho2017}
{Nugroho}, S.~K., {Kawahara}, H., {Masuda}, K., {et~al.} 2017, \aj, 154, 221

\bibitem[{{Parmentier} {et~al.}(2016){Parmentier}, {Fortney}, {Showman},
  {Morley}, \& {Marley}}]{Parmentier2016}
{Parmentier}, V., {Fortney}, J.~J., {Showman}, A.~P., {Morley}, C., \&
  {Marley}, M.~S. 2016, \apj, 828, 22

\bibitem[{{Parmentier} {et~al.}(2018){Parmentier}, {Line}, {Bean}, {Mansfield},
  {Kreidberg}, {Lupu}, {Visscher}, {D{\'e}sert}, {Fortney}, {Deleuil},
  {Arcangeli}, {Showman}, \& {Marley}}]{Parmentier2018}
{Parmentier}, V., {Line}, M.~R., {Bean}, J.~L., {et~al.} 2018, \aap, 617, A110

\bibitem[{{Parmentier} {et~al.}(2013){Parmentier}, {Showman}, \&
  {Lian}}]{Parmentier2013}
{Parmentier}, V., {Showman}, A.~P., \& {Lian}, Y. 2013, \aap, 558, A91

\bibitem[{{Rajpurohit} {et~al.}(2013){Rajpurohit}, {Reyl{\'e}}, {Allard},
  {Homeier}, {Schultheis}, {Bessell}, \& {Robin}}]{Rajpurohit2013}
{Rajpurohit}, A.~S., {Reyl{\'e}}, C., {Allard}, F., {et~al.} 2013, \aap, 556,
  A15

\bibitem[{{Rajpurohit} {et~al.}(2012){Rajpurohit}, {Reyl{\'e}}, {Schultheis},
  {Leinert}, {Allard}, {Homeier}, {Ratzka}, {Abraham}, {Moster}, {Witte}, \&
  {Ryde}}]{Rajpurohit2012a}
{Rajpurohit}, A.~S., {Reyl{\'e}}, C., {Schultheis}, M., {et~al.} 2012, \aap,
  545, A85

\bibitem[{{Sedaghati} {et~al.}(2017){Sedaghati}, {Boffin}, {MacDonald},
  {Gandhi}, {Madhusudhan}, {Gibson}, {Oshagh}, {Claret}, \&
  {Rauer}}]{Sedaghati2017}
{Sedaghati}, E., {Boffin}, H. M.~J., {MacDonald}, R.~J., {et~al.} 2017, \nat,
  549, 238

\bibitem[{{Sheppard} {et~al.}(2017){Sheppard}, {Mandell}, {Tamburo}, {Gand hi},
  {Pinhas}, {Madhusudhan}, \& {Deming}}]{Sheppard2017}
{Sheppard}, K.~B., {Mandell}, A.~M., {Tamburo}, P., {et~al.} 2017, \apjl, 850,
  L32

\bibitem[{{Southworth} {et~al.}(2007){Southworth}, {Wheatley}, \&
  {Sams}}]{Southworth2007}
{Southworth}, J., {Wheatley}, P.~J., \& {Sams}, G. 2007, \mnras, 379, L11

\bibitem[{{Spiegel} {et~al.}(2009){Spiegel}, {Silverio}, \&
  {Burrows}}]{Spiegel2009}
{Spiegel}, D.~S., {Silverio}, K., \& {Burrows}, A. 2009, \apj, 699, 1487

\bibitem[{{Stevenson} {et~al.}(2014){Stevenson}, {Bean}, {Madhusudhan}, \&
  {Harrington}}]{Stevenson2014}
{Stevenson}, K.~B., {Bean}, J.~L., {Madhusudhan}, N., \& {Harrington}, J. 2014,
  \apj, 791, 36

\bibitem[{{Swain} {et~al.}(2013){Swain}, {Deroo}, {Tinetti}, {Hollis},
  {Tessenyi}, {Line}, {Kawahara}, {Fujii}, {Showman}, \&
  {Yurchenko}}]{Swain2013}
{Swain}, M., {Deroo}, P., {Tinetti}, G., {et~al.} 2013, \icarus, 225, 432

\bibitem[{{Tinetti} {et~al.}(2007){Tinetti}, {Vidal-Madjar}, {Liang},
  {Beaulieu}, {Yung}, {Carey}, {Barber}, {Tennyson}, {Ribas}, {Allard},
  {Ballester}, {Sing}, \& {Selsis}}]{Tinetti2007}
{Tinetti}, G., {Vidal-Madjar}, A., {Liang}, M.-C., {et~al.} 2007, \nat, 448,
  169

\bibitem[{{Torres} {et~al.}(2012){Torres}, {Fischer}, {Sozzetti}, {Buchhave},
  {Winn}, {Holman}, \& {Carter}}]{Torres2012}
{Torres}, G., {Fischer}, D.~A., {Sozzetti}, A., {et~al.} 2012, \apj, 757, 161

\bibitem[{{Unsold}(1968)}]{Unsold1968}
{Unsold}, A. 1968, {Physik der Sternatmospharen, MIT besonder Berucksichtigung
  der Sonne}

\bibitem[{{Valenti} \& {Piskunov}(1996)}]{Valenti1996}
{Valenti}, J.~A. \& {Piskunov}, N. 1996, \aaps, 118, 595

\bibitem[{{Visscher} {et~al.}(2010){Visscher}, {Moses}, \&
  {Saslow}}]{Visscher2010}
{Visscher}, C., {Moses}, J.~I., \& {Saslow}, S.~A. 2010, \icarus, 209, 602

\bibitem[{{Zellem} {et~al.}(2017){Zellem}, {Swain}, {Roudier}, {Shkolnik},
  {Creech-Eakman}, {Ciardi}, {Line}, {Iyer}, {Bryden}, {Llama}, \&
  {Fahy}}]{Zellem2017}
{Zellem}, R.~T., {Swain}, M.~R., {Roudier}, G., {et~al.} 2017, \apj, 844, 27

\end{thebibliography}
\end{document}